\documentclass[sigconf,screen]{acmart}

\usepackage{xspace}

\usepackage{amsmath,amsfonts}
\usepackage{rotating}
\usepackage[symbol]{footmisc}
\usepackage{graphicx}
\usepackage{textcomp}
\usepackage{xcolor}
\usepackage{mathtools}
\usepackage{diagbox}
\usepackage{multirow}
\usepackage{pbox}
\usepackage{pdfpages}
\usepackage{titlesec }
\usepackage{algorithm} 
\usepackage{algorithmic}  
\usepackage[algo2e,ruled,vlined,linesnumbered]{algorithm2e}
\usepackage[tight,footnotesize]{subfigure}
\usepackage{xcolor, soul}
\usepackage[many]{tcolorbox}

\usepackage{enumerate}
\usepackage[shortlabels]{enumitem}
\usepackage[normalem]{ulem}

\usepackage{tcolorbox}

\newcommand*{\ie}{i.e.,\@\xspace}
\newcommand*{\eg}{e.g.,\@\xspace}

\newcommand*{\CB}{CodeBERT\@\xspace}
\newcommand*{\CO}{\textsc{Co3D}\@\xspace}

\definecolor{verylightgray}{gray}{0.99}
\definecolor{lightgray}{gray}{0.92}

\sethlcolor{lightgray}
\newcommand{\api}[1]{\texttt{\hl{\small #1}}}

\makeatletter
\newcommand*{\etc}{%
	\@ifnextchar{.}%
	{etc}%
	{etc.\@\xspace}%
}
\makeatother
\newcommand*{\etal}{\emph{et~al.}\@\xspace}
\newcommand\revised[1]{\textcolor{black}{#1}}

\newtcolorbox{shadedbox}{
	drop shadow southeast,
	breakable,
	enhanced jigsaw,
	colback=white,
	boxrule=0.80pt,
	left=0.3em,
	right=0.3em,
	top=0.1em,
	bottom=0.05em
}

\newcommand{\rqfirst}{\textbf{RQ$_1$}: \emph{Is the prediction obtained by C$_1$ and C$_2$ comparable to that by C$_3$?}} 
\newcommand{\rqsecond}{\textbf{RQ$_2$}: \emph{How does \CO compare with the baselines in term of prediction accuracy?}}
\newcommand{\rqthird}{\textbf{RQ$_3$}: \emph{Which technique is most timing efficient?}}

\usepackage{pifont}

\AtBeginDocument{%
  }

\setcopyright{acmcopyright}
\copyrightyear{2024}
\acmYear{2024}
\acmDOI{XXXXXXX.XXXXXXX}

\acmConference[EASE 2024]{The 28th International Conference on Evaluation and Assessment in Software Engineering}{18–21 June, 2024}{Salerno, Italy}

\begin{document}

\title{When simplicity meets effectiveness: Detecting code comments coherence with word embeddings and LSTM}


\author{Michael Dubem Igbomezie}
\small\affiliation{%
	\institution{Universit\`a degli studi dell'Aquila, Italy}
	\city{L'Aquila}
	\country{Italy}
}
\email{michaeldubem.igbomezie@student.univaq.it}


\author{Phuong T. Nguyen}
\small\affiliation{%
	\institution{Universit\`a degli studi dell'Aquila}
	\city{L'Aquila}
	\country{Italy}
}
\email{phuong.nguyen@univaq.it}

\author{Davide Di Ruscio}
\affiliation{%
        \institution{Universit\`a degli studi dell'Aquila}
	\city{L'Aquila}	
	\country{Italy}	
}
\email{davide.diruscio@univaq.it}

\begin{abstract}
	
Code comments play a crucial role in software development, as they provide programmers with practical information, allowing them to understand better the intent and semantics of the underpinning code. Nevertheless, developers tend to leave comments unchanged after updating the code, resulting in a discrepancy between the two artifacts. Such a discrepancy may trigger misunderstanding and confusion among developers, impeding various activities, including code comprehension and maintenance.  
Thus, it is crucial to identify if, given a code snippet, its corresponding comment is coherent and reflects well the intent behind the code. Unfortunately, existing approaches to this problem, 
while obtaining an encouraging performance, either rely on heavily pre-trained models, or treat input data as text, neglecting the intrinsic features contained in comments and code, including word order and synonyms.

This work presents \CO as a practical approach to the detection of code comment coherence. We 
pay attention to internal meaning of words and sequential order of words in text while predicting coherence in code-comment pairs. 
We deployed a combination of Gensim word2vec encoding and a simple recurrent neural network, a combination of Gensim word2vec encoding and an LSTM model, and CodeBERT. 
The experimental results show that \CO obtains a promising prediction performance, thus outperforming well-established baselines. \revised{We conclude that depending on the context, using a simple architecture can introduce a satisfying prediction.}

\end{abstract}

\maketitle

\section{Introduction}
\label{sec:Introduction}

Textual comments are embedded in source code as an effective means to facilitate the readability of the code. Essentially, comments are valuable pieces of information when it comes to code understanding and maintenance~\cite{DBLP:journals/ese/PascarellaBB19,DBLP:conf/kbse/WeiLC16}, revealing various aspects of the underlying method
~\cite{DBLP:journals/ese/CasseeZNSP22,7332619,DBLP:conf/msr/PerumaANMO22}. 
Among other purposes, developers use comments 
to communicate their intent and 
explanations about the implementations. This helps developers spend less time understanding the existing implemented solutions. However, as pointed out by Corazza \etal \cite{DBLP:conf/clic-it/CorazzaMS16} \textit{``information provided in the comment of a method and in its corresponding implementation may not be coherent with each other (\ie the comment does not properly describe the implementation).''}
Software evolution~\cite{10.1145/3530786}, among others, is 
a factor that could break the coherence between a comment and the block of code it originally intended to describe. 

\begin{figure*}[t!]
	\centering
	\begin{tabular}{cc}	
		\subfigure[Coherent]{\label{fig:Coherent}
			\includegraphics[width=0.46\linewidth]{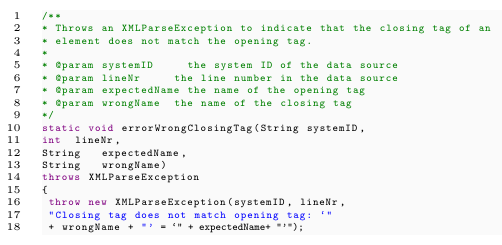}} &		
		\subfigure[Incoherent]{\label{fig:Incoherent}
			\includegraphics[width=0.48\linewidth]{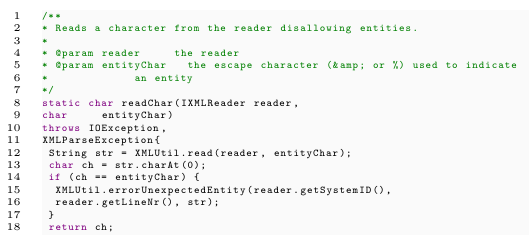}} 	
	\end{tabular} 
	\vspace{-.2cm}
	\caption{Examples of coherent and incoherent code comment pairs (Extracted from the dataset of Corazza \etal \cite{DBLP:journals/sqj/CorazzaMS18}).} 	
	\label{fig:Examples}
	\vspace{-.4cm}
\end{figure*}

As a motivating example, we depict two code snippets\footnote{Both snippets are extracted from the dataset curated by Corazza \etal \cite{DBLP:journals/sqj/CorazzaMS18}.} with comments in Fig.~\ref{fig:Examples}. The first snippet (see Fig.~\ref{fig:Coherent}) throws an \api{XMLParseException} if the opening and closing tags do not match. By checking the comment, we realize that the intent of the code is well explained, corresponding to a \emph{coherent} relationship between code and comment. Meanwhile, the code in Fig.~\ref{fig:Incoherent} is used to check if the first character of the input string matches with the one specified by the \api{entityChar} variable, and if this is the case, then an error will be thrown. Looking at the comment, we see that it does not properly explain the behaviour of the code. Thus, we identify an \emph{incoherent} case, \ie the comment implies a completely different story with respect to the one contained in the code. Such inconsistency may cause misunderstanding, leading to various issues, including difficulties in software maintenance. 


Identifying instances where there is a lack of coherence between code methods and their accompanying comments could significantly improve software practice, and reduce the time spent in source code integration. Corazza \etal \cite{DBLP:journals/sqj/CorazzaMS18} were among the first to address the issue, and they curated a dataset where code and comment were manually selected and evaluated. Since then, there have been various studies proposed to the detection of coherence between comments and code~\cite{DBLP:conf/iwpc/RabbiS20,DBLP:journals/corr/abs-2207-14444}. Nevertheless, while achieving encouraging prediction performance, existing approaches encode code and comments using conventional encoding models, ignoring the intrinsic features contained in each individual artifact. 



\revised{The proliferation of pre-trained and large language models (LLMs) in recent months has transformed Software Engineering (SE) research~\cite{DBLP:journals/software/Ozkaya23b}. Various studies have used pre-trained models and LLMs as a means to gear their performance, and in fact LLMs have demonstrated their applicability in different SE tasks. At a certain point, the following question arises: Will the traditional machine learning algorithms become obsolete because of such models? We hypothesize that though heavily pre-trained and large models, while being trendy, they are not \emph{a silver bullet}, \ie a solution to any problems. These models are resource-demanding, and training or fine-tuning them necessitates computing power and, thus, energy. Moreover, depending on the context of the application, we can obtain a satisfying 
	performance without resorting to an overkill machine learning model.} 



In this paper, we propose \CO--a workable solution to the detection of \textbf{Co}de \textbf{Co}mment \textbf{Co}herence \textbf{D}etector. 
Instead of using a complex 
model as the classification engine, we conceived a tool 
on top of three different techniques as follows: 
\textbf{C$_1$}: a combination of Gensim word2vec 
and a simple recurrent neural network; \textbf{C$_2$}: a combination of Gensim word2vec and a long short-term memory neural network (LSTM)~\cite{DBLP:journals/neco/HochreiterS97}; 
and \textbf{C$_3$}: tokenization and \CB~\cite{feng2020codebert}. %
\revised{The first two methods can be considered a manual version of the last two methods, with the last two methods involving 
the use of transfer learning from large pre-trained models.} 
The aim is to shed light on what exactly these large pre-trained models do differently from the traditional and older methods. The \CB pre-trained model is both used as a means to check whether the corpus used in encoding texts play any significant role in better prediction. 
While 
C$_1$ and C$_2$ are conceived to capture the internal meaning, 
as well as sequential ordering of words, 
C$_3$ is built on top of a pre-trained model. 
We came across interesting results, \ie despite being simple architectures, C$_1$ and C$_2$ bring an encouraging 
performance, compared to using BERT and \CB--two more complicated infrastructures. 
It is important to highlight that the proposed approach does not necessitate a pre-training phase, distinguishing it from the considered baselines. Despite its straightforward nature, \CO outperforms two well-founded 
baselines, providing evidence for the need to explore the application of these technologies in addressing the challenge. 

The main contributions of our work are summarized as follows: 

\begin{itemize}
	\item We propose \CO--a practical approach to the detection of code comment coherence using word embedding, LSTM, and \CB. Thus,  \revised{we challenge the existing implicit assumptions 
	in the field, \ie pre-trained models and LLMs are supposed to be the silver bullet for solving all possible issues in software engineering.} 
	\item  While we defer a rigorous comparison of \CO with LLMs as future work, we evaluate our approach against well-defined pre-trained models including BERT and CodeBERT using a real-world dataset. 
	\item The tool developed through this paper together with metadata is published to allow for future research.\footnote{\url{https://github.com/EASE2024-Co3D/Co3D/}}
\end{itemize}

\section{The \CO approach}
\label{sec:Methodology}

%
\revised{This section describes our \CO approach to code comment coherence classification.} 

\subsection{Motivation}





\textbf{Synonym.} The internal meaning of words makes it possible for two sentences--made up of entirely of synonyms
--to be interpreted as having similar or same intent, \eg 
``\emph{The baby dog crawled below the fence,}'' and ``\emph{The puppy creeped under the wall,}'' despite not containing 
the exact same words, should be interpreted as having the same information or intent. \textbf{Word order.} The 
order of words in a sentence 
differentiates two sentences made up entirely of the same words, but with different orders. For example, ``\emph{The student listens to the teacher,}'' and ``\emph{The teacher listens to the student,}'' have 
different meanings, though they share the same set of words. Attention to the first kind of meaning explained could play a significant role in developing a model that generalizes beyond the exact words in the dataset with which it was trained. While attention to the second kind of embedded meaning could be significant in a given instance where there are two methods being inverse of each other, where the first method takes the output of the second method as input and produces an output which can serve as the input to the second method. If both methods are documented with the same comment, considering the second meaning explained, this would be a wrong comment. However, this inaccuracy might not be recognized by a classification method which does not pay attention to the sequential structure of sentences.

In code comment, we encounter a similar phenomenon. %
Consider a method that takes base2 values as input, and produces base10 values, then the method is associated with the following comment: \api{/*Takes\; base2\; values\; as\; input\; and\; outputs\; base10\; values*/}. 
There is the second method that takes base10 values as input, and outputs base2 values, however it has the following comment \api{/*Takes\; base2\; values\; as\; input\; and\; outputs\; base10\; values*/}. 
An encoding method 
that does not take into account word order, is not able to recognize that the second code-comment pair is actually incoherent, as the correct version should be \api{/*Takes\; base10\; values\; as\; input\; and\; outputs\; base2\; values*/}. 


%


\begin{figure}[h!]
	\centering
	\includegraphics[width=0.45\textwidth]{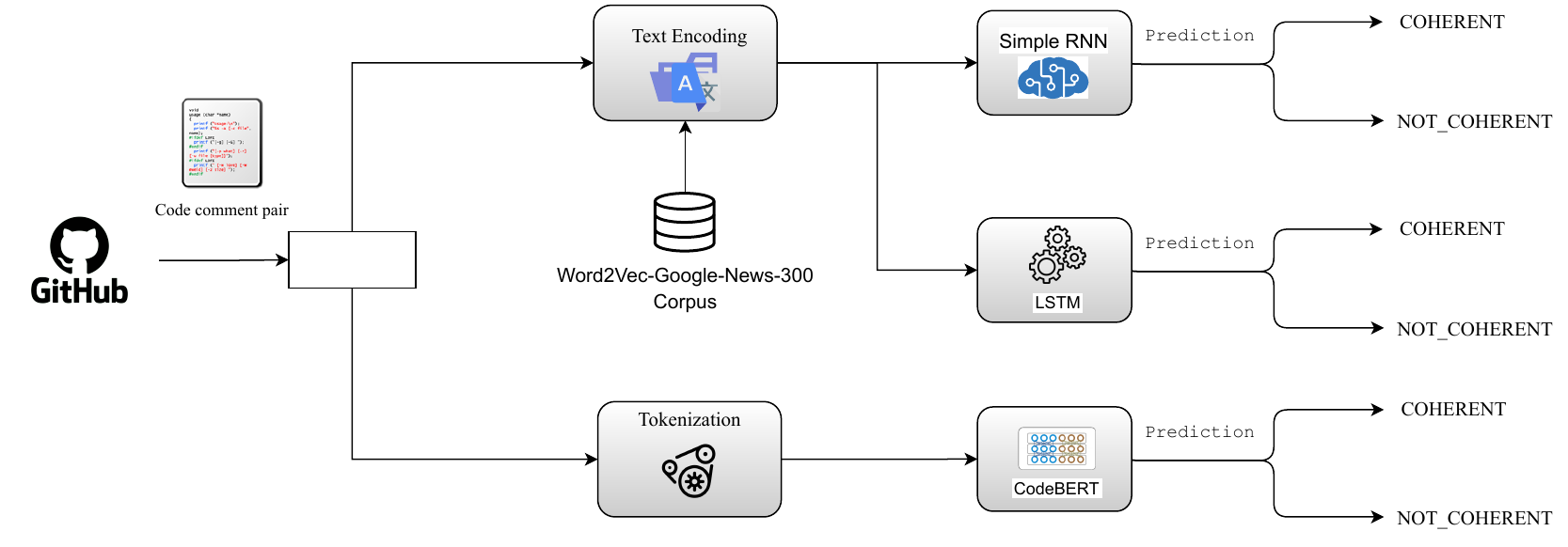}
	\caption{The overall architecture.}
	\label{fig:Architecture}
	\vspace{-.2cm}
\end{figure}

\subsection{Classification Engine}

We address the aforementioned issues 
by focusing on the 
encoding method used on the text accounts for the internal meaning of words, and the classification algorithms to account for the sequential word order. 
This is done by using a combination of the Gensim word2vec encoding method \cite{DBLP:journals/corr/abs-1301-3781}, 
and a Simple RNN \cite{JORDAN1997471}, or an LSTM model \cite{DBLP:journals/neco/HochreiterS97}. 
To validate the efficiency, we also use \CB together with tokenization to encode an entire sentence, paying attention to both the internal meaning and sequential word order. 

Simple RNN and LSTM are specified with the same configuration, \ie both having a single hidden layer with 100 nodes, and an output layer with dense connections. The categorical cross entropy is used to compute losses, the Adam optimizer is for defining the learning rate, with a batch size of 50, and both models were trained using back propagation. 
The models were trained using two Tesla T4 GPUs. 
The Adam optimizer was applied in the training of \CB. We split the entire dataset following the 0.8:0.2 ratio, \ie 80\% and 20\% of the data are used for training and testing, respectively. In Fig.~\ref{fig:Architecture}, we depict the proposed \CO architecture. 
The three configurations are as follows: \textbf{C$_1$}: Simple RNN + Gensim word2vec; \textbf{C$_2$}: LSTM + Gensim word2vec; \textbf{C$_3$}: Tokenization + \CB.

\subsection{Data Converting}

 
The Gensim \api{word2vec-google-news-300} corpus\footnote{\url{https://huggingface.co/fse/word2vec-google-news-300}} is utilized to encode data for Simple RNN and LSTM. In particular, each code-comment pair is tokenized, whereby a token is represented 
using a vector of 300 entries. Since the length of sentences varies, each entry of the list of vectors 
is constrained to a maximum length of 50 vectors. Such a number is not arbitrarily chosen, as the average word length of all code-comment pairs was 113 words. However, due to the polynomial time complexity, parsing through each token of each entry is time demanding, so this average length is approximately halved (50), and it is then used to truncate longer entries or pad shorter entries. After this step, the input data is then converted into a tensor of size $n$ $\times$ 50 $\times$ 300 ($n$ the is number of rows), which is eventually fed as input for Simple RNN and LSTM.

\section{Evaluation}
\label{sec:Evaluation}
This section introduces the research objectives, and 
the evaluation conducted 
using an existing dataset and two baselines.

\subsection{Research Questions} \label{sec:ResearchQuestions}


We study \CO with 
the following research questions:


\begin{itemize}
	\item \rqfirst~C$_1$, C$_2$ are the two configurations taking into account word meaning and order when learning code comment pairs, while C$_3$ is built on top of a pre-trained model. We study if lightweight models also yield a comparable accuracy with respect to more complex ones. 
	\item \rqsecond~By using the same datasets and experimental settings, we compare \CO with two state-of-the-art approaches~\cite{DBLP:journals/sqj/CorazzaMS18,DBLP:journals/corr/abs-2207-14444} in the detection of code comment coherence that work on top of SVM and BERT--two state-of-the-art approaches in the detection of code comment coherence. 
	\item \revised{\rqthird~We are interested in understanding the timing efficiency of the considered techniques with respect to text embedding, and conversion time. This is an important aspect as prolonged processing time may make a model less attractive in practice.}
\end{itemize}

\subsection{Dataset, baselines, and metrics}


We use the dataset curated by Corazza \etal \cite{DBLP:journals/sqj/CorazzaMS18}, 
containing code-comment pairs and their coherence status from the Benchmark, CoffeeMaker, JFreeCha\-rt060, JFreeChart071 and JHotDraw741 methods \cite{DBLP:conf/clic-it/CorazzaMS16}, which possess 2,881, 47, 461, 588, and 1,785 code-comment pairs, respectively. 
The dataset was originally in the \api{.txt} format were all parsed through cleaning and transformation functions, and converted into .csv files so as to be easily read into the Jupyter environment using pandas.

\begin{table}[h!]
	\centering
	\vspace{-.1cm}
	\caption{Dataset Size.}  
	\label{tab:Datasets}
	\begin{tabular}{|p{2.6cm}|p{2.5cm}|}  \hline  
		& Number of Rows  \\ \hline		
		Benchmark & 2,881   \\ \hline            
		CoffeeMaker & 47  \\ \hline
		JFreeChart060 & 461  \\ \hline
		JFreeChart071 & 588  \\ \hline
		JHotDraw741 & 1,785  \\ \hline            
		All & 5,762  \\ \hline           
	\end{tabular}	
	\vspace{-.1cm}
\end{table}

Concerning the baselines, we opt for the work by Corazza \etal \cite{DBLP:conf/clic-it/CorazzaMS16}, which employed Bag of Word (BoW) to encode input data, and SVM 
as the classification engine. Moreover, we consider a more recent tool developed by Steiner and Zhang~\cite{DBLP:journals/corr/abs-2207-14444} as another baseline, built of top of BERT \cite{DBLP:conf/naacl/DevlinCLT19}. 
The encoding for BERT is very similar to that of the Gensim corpus. The input data is parsed through the BERT-based tokenizer, where 
each word is converted into its corresponding vector, then the entire sequence of vectors is encoded as a sentence. After this, the special tokens necessary for fine tuning BERT are added to the encoded sentence, finally the entire vector representation of the sentence is either truncated or padded to the chosen maximum length. 
To aim for a fair comparison, we run both baselines using their original implementation on the same dataset. 

For the evaluation metrics, we make use of Accuracy, Precision, Recall, and F$_1$ score~\cite{Dalianis2018} as they have been widely used in evaluating machine learning systems. Due to the space limit, we cannot introduce the metrics here, interested readers are kindly referred to a chapter by Dalianis \cite{Dalianis2018} for more detail.


%
%
%


%
%
%
%
%


\section{Experimental Results}
\label{sec:Results}

Table~\ref{tab:Results} reports the evaluation metrics where the best values are printed in bold. After training the models 
on each of the datasets, and then training them on the combined datasets (the rows with ``\textbf{All}''), we obtained the scores which are then averaged out. 
We analyze the results by answering the research questions (see Section~\ref{sec:ResearchQuestions}) as follows.  

\begin{table}[t!]
	\centering
	\small
		\vspace{-.1cm}
	\caption{Evaluation results.}  
		\label{tab:Results} 
	\begin{tabular}{|p{0.3cm}|p{1.6cm}|p{0.80cm}|p{0.70cm}|p{0.70cm}|p{0.70cm}|p{0.85cm}|}  \hline  
		& & \multicolumn{3}{c|}{\CO} & \multicolumn{2}{c|}{Baselines} \\ \cline{2-7} 
		& \textbf{Dataset} & C$_1$ & C$_2$  & C$_3$ & SVM & BERT  \\ \hline 
		
		 	
       {\multirow{6}{*}{\rotatebox[origin=c]{90}{Accuracy}}} &    Benchmark & 0.790 & 0.800 & \textbf{0.847} & 0.826 & 0.813  \\ \cline{2-7}
            
        &    CoffeeMaker & \textbf{1.000} & 0.958 & 0.917 & 0.778 & 0.875  \\ \cline{2-7}

        &    JFreeChart060 & 0.916 & \textbf{0.921} & 0.875 & 0.876 & 0.883  \\ \cline{2-7}

        &    JFreeChart071 & 0.909 & \textbf{0.910} & 0.878 & 0.875 & 0.871  \\ \cline{2-7}

        &    JHotDraw741 & 0.752 & 0.761 & 0.787 & \textbf{0.805} &  0.769  \\ \cline{2-7}
            
        &    All & 0.924 & 0.928 & \textbf{0.929} & 0.928 & 0.879  \\ \hline
            
            
       {\multirow{6}{*}{\rotatebox[origin=c]{90}{Precision}}}  &    Benchmark & 0.813 & 0.818 & \textbf{0.836} & 0.830 & 0.798  \\ \cline{2-7}
            
        &    CoffeeMaker & \textbf{1.000} & \textbf{1.000} & 0.833 & 0.724 & 0.944  \\ \cline{2-7}
            
        &    JFreeChart060 & 0.940 & \textbf{0.941} & 0.908 & 0.917 & 0.883  \\ \cline{2-7}
            
        &    JFreeChart071 & \textbf{0.931} & 0.921 & 0.878 & 0.908 & 0.878  \\ \cline{2-7}
            
        &    JHotDraw741 & 0.704 & 0.724 & \textbf{0.787} & 0.774 & 0.689  \\ \cline{2-7}
            
        &    All & 0.932 & \textbf{0.941} & 0.920 & 0.935 &  0.901  \\ \hline

            
        {\multirow{6}{*}{\rotatebox[origin=c]{90}{Recall}}} &   Benchmark & 0.848 & 0.862 & 0.846 & \textbf{0.885} & 0.823  \\ \cline{2-7}
            
        &    CoffeeMaker & \textbf{1.000} & 0.937 & \textbf{1.000} & 0.952 & 0.833  \\ \cline{2-7}
            
        &    JFreeChart060 & 0.970 & 0.975 & 0.954 & 0.944 & \textbf{1.000}  \\ \cline{2-7}
            
        &    JFreeChart071 & 0.972 & 0.986 & \textbf{1.000} & 0.957 & \textbf{1.000}  \\ \cline{2-7}
            
        &    JHotDraw741 & 0.745 & 0.749 & 0.785 & 0.762  & \textbf{0.836} \\ \cline{2-7}
            
        &    All & \textbf{0.968} & 0.963 & 0.928 & 0.954  & 0.851  \\ \hline

            
        {\multirow{6}{*}{\rotatebox[origin=c]{90}{F$_1$ score}}} &     Benchmark & 0.830 & \textbf{0.839} & 0.836 & 0.857 & 0.807  \\ \cline{2-7}
            
        &    CoffeeMaker & \textbf{1.000} & 0.966 & 0.905 & 0.814 & 0.867  \\ \cline{2-7}
            
        &    JFreeChart060 & 0.955 & \textbf{0.958} & 0.928 & 0.930 & 0.938  \\ \cline{2-7}
            
        &    JFreeChart071 & 0.951 & \textbf{0.952} & 0.933 & 0.932 & 0.933  \\ \cline{2-7}
            
        &    JHotDraw741 & 0.724 & 0.731 & \textbf{0.785} & 0.768 & 0.742  \\ \cline{2-7}
            
        &    All & 0.950 & \textbf{0.952} & 0.922 & 0.944 & 0.868  \\ \hline           
            
	\end{tabular}	
	\vspace{-.4cm}
\end{table}

\subsection{\rqfirst}

By comparing the accuracy scores by C$_1$ and C$_2$ with C$_3$ in the top of Table~\ref{tab:Results}, we can see that using word2vec together with Simple RNN and LSTM brings a comparable performance in relation to using \CB. In particular, by C$_1$ and C$_2$, \CO obtains the best accuracy by CoffeeMaker, JFreeChart060, and JFreeChart071; While by C$_3$, it gets the maximum accuracy with Benchmark and All. 
With respect to the Precision scores, we witness a similar trend: using simple architectures built on top of word2vec combined with Simple RNN and LSTM allows \CO to obtain a similar, and somehow better performance compared to employing \CB. Especially, \CO yields a maximum Precision of 1.000 for both C$_1$ and C$_2$ by the CoffeeMaker category. 
By the Recall scores, we also see that C$_1$ and C$_2$ bring a better performance compared to C$_3$. This is further confirmed when we consider the F$_1$ scores as the final evaluation metric, \ie C$_1$ and C$_2$ account for four best scores among six categories, while C$_3$ does this by the remaining two categories. 

\revised{The results in this research question show that even with a simple architecture, we can obtain a satisfying prediction performance, compared to complex techniques, \ie \CB.}

\begin{shadedbox}
	\small{\textbf{Answer to RQ$_1$.} Using two simple architectures, \ie Gensim word2vec combined with Simple RNN and LSTM, we obtain a satisfying prediction performance, compared to using \CB--a more complex pre-trained model.}
\end{shadedbox}

\subsection{\rqsecond}



In this research question, we evaluate \CO by its three configurations, \ie C$_1$, C$_2$, and C$_3$, against the two baselines including Corazza \etal \cite{DBLP:journals/sqj/CorazzaMS18} built of top of SVM, and Steiner and Zhang~\cite{DBLP:journals/corr/abs-2207-14444} with the BERT classification engine. Following Table~\ref{tab:Results}, it is evident that \CO 
outperforms the SVM baseline model by all the four considered metrics. In fact, SVM only gets the maximum scores by Accuracy and Precision with the JHotDraw741 and Benchmark categories, respectively. Apart from that, \CO obtains superior scores by almost all the remaining rows. Among others, C$_2$ brings more highest scores for \CO by different data categories.


This same conclusion even becomes somewhat clearer when the best accuracy of all models are compared in a similar way. Comparing the models proposed in this paper using the same metrics, the CodeBERT model combined with the CodeBERT tokenizer comes out on top, followed by the Simple RNN and then the LSTM model, with slight differences in their performance. It is also important to note the loss of information during the data transformation for the Simple RNN and LSTM models could also be a contributing factor to their current performance, and could improve significantly if the maximum length of words/elements in each code-comment pair encoded is increased to match the average number of words/elements per code-comment pair.

Similarly, compared to BERT, \CO also yields better evaluation scores by almost all the rows in Table~\ref{tab:Results}. In particular, among the total 24 rows, BERT gets maximum Recall scores by three rows, \ie for JFreeChart060, JFreeChart071, and JHotDraw741. Meanwhile, \CO yields maximum scores by 19 over the 21 remaining rows across all the three configurations, \ie C$_1$, C$_2$, and C$_3$.



%

\revised{In summary, we conclude that the model proposed in this paper outperforms the considered baselines \cite{DBLP:journals/sqj/CorazzaMS18,DBLP:journals/corr/abs-2207-14444}. Complementing these models with each of their strengths, like the level of information retained by the input to the RNN models, the corpus used to tokenize and train the CodeBERT, the size and functioning principle of the CodeBERT model (like the Bi-directional attention to sequence order of words in text \cite{feng2020codebert,DBLP:journals/corr/VaswaniSPUJGKP17}), it is possible to achieve a higher accuracy in future research.}

\begin{shadedbox}
	\small{\textbf{Answer to RQ$_2$.} On the considered dataset, with the three internal configurations, \CO outperforms both baselines with respect to Accuracy, Precision, Recall, and F$_1$ score.}
\end{shadedbox}

\subsection{\rqthird}

Table~\ref{tab:EmbeddingTime} shows the text embedding time (in minutes) for each of the text embedding methods used by each of the classification engines. The recorded time shows the duration it takes for the embedding method used by each of the classification engines to convert the raw text into vector format before training. From the record, we can observe that the tokenizer used by \CB yields the fastest conversion time amongst all the tokenizers. The TF-IDF embedding schema used by the SVM baseline was the second fastest, closely followed by the BERT Tokenizer. The Word2Vec embedding method has the worst conversion time, and far outweighs the conversion of the rest of the methods explored (even when combined). This could depend on the nature and size of the embedding model used in the Word2Vec method, as well as the fact that each individual word in each code-comment pair is searched and converted separately as opposed to the sentence embedding used in the likes of BERT and CodeBERT. The TD-IDF embedding method is also relatively simple compared to the Word2Vec method.

\begin{table}[h!]
	\centering
		\vspace{-.1cm}
	\caption{Text Embedding Time (in minutes).}  
		\label{tab:EmbeddingTime}
	\begin{tabular}{|p{1.6cm}|p{1.2cm}|p{1.3cm}|p{1.0cm}|p{1.0cm}|p{1.0cm}|}  \hline  
		& \multicolumn{2}{c|}{\CO} & \multicolumn{2}{c|}{Baselines} \\ \hline 
		& Word2Vec & CodeBERT  & BERT & TF-IDF  \\ \hline
		
           Benchmark & 50 & \textbf{0.017} & 0.167 & 0.133  \\ \hline
            
           CoffeeMaker & 0.80 & \textbf{0.001} & 0.001 & 0.001  \\ \hline

           JFreeChart060 & 7.0 & \textbf{0.005} & 0.017 & 0.006  \\ \hline

           JFreeChart071 & 9.0 & \textbf{0.006} & 0.031 & 0.007  \\ \hline

           JHotDraw741 & 32.0 & \textbf{0.020} & 0.100 & 0.060  \\ \hline
            
           All & 98 & \textbf{0.067} & 0.333 & 0.250  \\ \hline
           
	\end{tabular}	
	\vspace{-.1cm}
\end{table}



Considering these numbers, as well as the model performance from the various classification engines explored in this research, \CB is the best model in terms of both performance and execution time, despite the fact that the SimpleRNN and LSTM were able to match the CodeBERT method in performance. If these two methods were to be employed in a real-time situation, some modification might be needed to mitigate this significantly longer conversion time. Some of these methods could involve the integration of distributed processing or the use of an entirely different embedding method along with the Simple RNN and LSTM method.

\begin{shadedbox}
	\small{\textbf{Answer to RQ$_3$.} Among the considered models, \CB is the most timing efficient one.}
\end{shadedbox}

\section{Discussion}
\label{sec:Discussion}


\revised{In this section, we interpret the results obtained from the experiments, as well as discuss the implications of our findings. Moreover, the threats to validity are also highlighted.} 

\subsection{Model Interpretation} \label{sec:Interpretation}

Considering the further integration of AI into much of our daily activities and online interactions, there is a rising need to explain these models for ethical and legal reasons, as well as for general acceptability. The public is more likely to accept these integrations when they can look below the hood to get an idea of what principles drive them. 
It can also serve as a means to confirm the underlying theories that lead up to selection of these models and how they are expected to work.

For this purpose, the $transformers{\_}interpret$ library was used to parse a sample code-comment block along with the trained CodeBERT model and the importance of each token towards the final predicted class of the sample was visualized.

\begin{figure}[h!]
	\centering
	\includegraphics[width=0.5\textwidth]{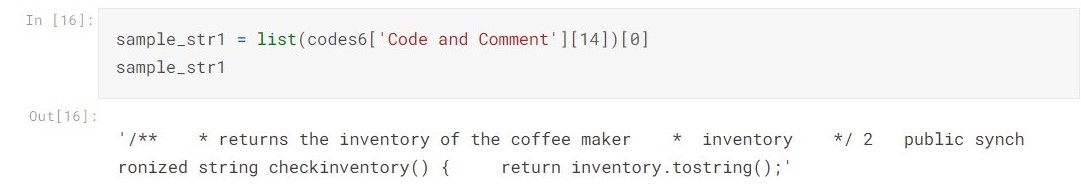}
	\caption{A selected sample code-comment pair used for the sake of model interpretation.}
	\label{fig:SampleCode-Comment Pair}
	\vspace{-.2cm}
\end{figure}

\begin{figure}[h!]
	\centering
	\includegraphics[width=0.45\textwidth]{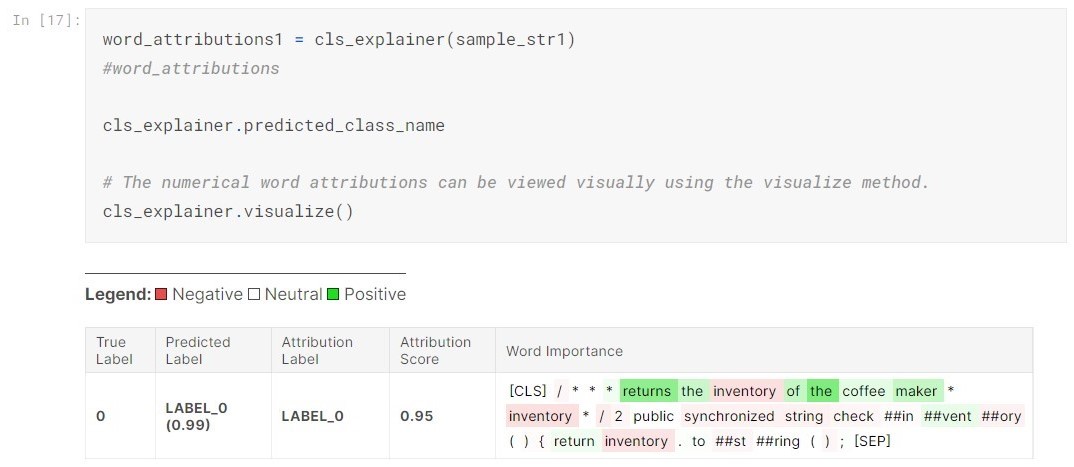}
	\caption{Visualization of the importance of each word in the final predicted class.}
	\label{fig:Model Interpretation}
	\vspace{-.2cm}
\end{figure}

To give a high-level explanation of the visualization in Figure \ref{fig:Model Interpretation}, first we start by explaining the significance of the colour coding used by the library to render each text in the sample code-comment pair. Similar to how colour codes are used in a statistical correlation heat-map, the green colour indicates positive direction and the red colour indicates a negative direction, while the intensity of the colour used to render a particular token shows the level of impact that particular token has in the positive or negative direction. This can also be likened to the hypothesis check conducted on coefficients of a given regression model to obtain their p-value and conclude whether they can be considered statistically significant or not. Simply put, this can be seen as the positive or negative relationship between the input and the output, and the importance of this relationship. 

Taking the above explanation into consideration and observing the results of the visualization, we can see the words like $return$, $the$, and $maker$ having the most positive relationship, while the word $inventory$ has the most negative relationship, but all come together to aid the model in deciding the given category for this particular sample. Also as expected of the intent behind the methods used in this research work (accounting for semantic meaning and sequence order), we can observe how similar words appearing later down the line helped in correctly classifying this sample. We can see this in how the word $return$ reoccurs, having positive relations both times. This same effect occurs with the word $inventory$. When this input was being passed through the model, the model was encouraged to pay more attention to these words, due to their recurrence in both the comment and the code. This also makes sense when judging this from a programmer's point of view, as these set of words actually help show that the code-comment pair is coherent. Regardless of the fact the the middle part of the sample containing ``${public\ synchronized\ string}$'' is still very important in the functionality of the code, the words prioritized by the model give the sample code-comment pair some context in terms of checking what the comment stated the code will do, and what the code is eventually doing.

\subsection{Implications}

The outcomes of our empirical evaluation reveal an interesting finding: Even with simple machine learning (ML) architectures, we can obtain a satisfying performance compared to that by pre-trained models, \ie BERT and \CB. \revised{The experimental results show that \CO outperforms both BERT and SVM.} This is \emph{perversely counter-intuitive}, as we expect that these models--being trained on large corpora of source code and text--such as BERT or \CB should have been the best classifier, surprisingly they are not. 
This is important in practice as training on complex pre-trained models 
requires time and computational resources, and thus using lightweight models while still preserving a comparable accuracy is highly beneficial. 
Apart from technical consequences, this has an obvious social impact, as we can avoid spending too much energy (\eg electricity) to train complex deep learning models, instead we just employ a simple, yet efficient and effective ML technique.


One of the main differences between BERT and \CB is the 
corpora for encoding and training. The corpus for \CB is more suitable for input text involving code and programming languages \cite{feng2020codebert}, implying that some tokenized words might have a better representation after being encoded, which will in turn lead to better pattern recognition during training or fine tuning. Through the experiments, we see that although BERT and \CB have been successful in various tasks~\cite{DBLP:conf/dexa/BatraPSA21,9808712,ISLAM2023119919}, they are not a \emph{silver bullet}, \ie a solution to every problem. 
Altogether, this calls for fundamentally new directions of future research, where we need to empirically select 
a model that fits to a specific problem, rather than picking a model for any probable Software Engineering issue, just because the model is either `\emph{advanced}' or `\emph{deep}.'


A major difference between the output of this encoding method and the output of the encoding method used for the Simple RNN and LSTM is that unlike the latter, the former does not represent each word as its own vector, but represents the entire sentence as a single vector of the specified maximum length. One could argue that less information is retained due to this difference, which begs the question, “If this caused some significant loss of information, could it be compensated for by the size of the BERT and CodeBERT models?” All these were the various curiosities sought to be addressed by this paper.


\subsection{Threats to validity}

\textbf{Internal validity.} 
The threat is related to both data quantity \cite{4804817} and quality, and this can be seen from the overall model performance on the JHotDraw741 dataset, which 
is with the lowest Kappa index, \ie 
below 0.70 \cite{DBLP:conf/clic-it/CorazzaMS16} 
across all the experimented models. 

Another notable threat is the amount of information discarded during the text encoding section of the research. 
Each code-comment pair was truncated or padded to the first 50 vectors corresponding to the first 50 word/symbols from each code-comment pair (for the Simple RNN and LSTM pairs). This can be a loss of information considering that the longest code-comment pair had a word/symbol length above 500. \revised{Similarly for the BERT and \CB model (although less extreme), each entry was truncated to average length of 113 which is still significantly smaller compared to the max recorded length above 500, and the maximum allowable length of both models being above 700 tokens.} 
We anticipate that with a more distributed and parallel approach to this encoding process, 
the models might produce better predictions than they already did.

\textbf{External validity.} This concerns the generalizability of our approach. 
Only the Java programming language was considered for the evaluation. 
Therefore, we do not know if the results obtained are specific to Java, or if it will be applicable to other languages. As we have already seen from this paper, the corpus (which can be language specific) does seem to play a role in model performance, hence can also be considered a threat to the work. Lastly, a more generic threat to this paper is the size of the dataset, \ie more data is supposed to bring better training and inference
 \cite{4804817}. 

\section{Related Work}
\label{sec:RelatedWork}
The first work to address the issue of code comment coherence was %
published by Corazza \etal \cite{DBLP:journals/sqj/CorazzaMS18}. 
Three different approaches were used to investigate code-comment coherence: \emph{(i)} Checking if the lexical similarity between the code and comment calculated using cosine similarity had any correlation with the coherence of each pair; \emph{(ii)} Executing a dimensionality reduction method (PCA) on the vector encoded dataset using the vector space model (VSM), and visualizing the result to check if there is some obvious divide between coherent and non-coherent pairs; and \emph{(iii)} Using support vector machines (SVM) to train and predict on the encoded dataset.


\revised{Similar research was conducted \cite{DBLP:conf/iwpc/RabbiS20}, encoding data using abstract syntax tree of the source code to create vocabulary for tokenizing the dataset instances and 
feeding these tokens into a siamese neural network and an LSTM. The purpose of the second work was to account for the inner meaning of code and comments and the sequence order.} 


A recent approach~\cite{DBLP:conf/seke/HaqueKSKR20} employed ensemble learning to detect code-comment coherence 
to improve on the vector space encoding approach used in the first paper by Corazza \etal \cite{DBLP:journals/sqj/CorazzaMS18}. The authors used Latent Dirichlet Allocation (LDA) to summarize similar words and represented them with the same topic. This was then combined with an ensemble approach (Random Forest) to replace SVM used by Corazza \etal \cite{DBLP:journals/sqj/CorazzaMS18}. Another work \cite{DBLP:conf/iwpc/RabbiS20} tried to improve Corazza \etal \cite{DBLP:journals/sqj/CorazzaMS18} by capturing the sequential order in words. 
The major difference with our work lies in the encoding method, which was still limited to the dataset used to train the model, that does not necessarily capture the internal or synonyms of these words, 
this could limit the generalization of the model to only words encountered in the dataset. 

The most recent work similar to ours is the one by Steiner and Zhang \cite{DBLP:journals/corr/abs-2207-14444}, which considered the use of large pre-trained models, \ie BERT and Longformer, for code-comment coherence check. However, unlike the aforementioned approach, 
our work focuses on an in-depth look into the factors that differentiate these pre-trained models from the traditional ones, \revised{like the SVM model used by Corazza et al \cite{DBLP:journals/sqj/CorazzaMS18}, 
and whether some of the factors of these large pre-trained models were necessary for the subject matter.} 
Large pre-trained models 
can be limited in their customizability, so approaches employed these pre-trained models could be replicated using smaller and more customizable models, while achieving a similar or even higher performance. This is the case with our approach, in this work we show that even simple architectures can have a better performance compared to the approach built on heavily pre-trained models 
\cite{DBLP:journals/corr/abs-2207-14444}. 

\section{Conclusion}
\label{sec:Concluion}
In this paper, we performed an investigation into
the contributing factors of the NLP implementation in code-comment coherence prediction. 
Three models were trained 
and some included transfer learning where large pre-trained models were fine-tuned on the available dataset, and the results were compared to some baseline models from the original paper 
\cite{DBLP:journals/sqj/CorazzaMS18}. We conclude that even simple architectures are both effective and efficient compared to complex pre-trained models.
\revised{All results obtained from the experiment points to the confirmation of this suspicion, that these two factors contribute to improving the models' performance. Additionally it was observed that the size of models did not seem to matter after these two factors have been successfully accounted for, at least with the current parameter and configuration of the experiment. Finally it was also observed that the nature of the corpus used for the encoding process also plays a significant role in predicting code-comment coherence.} This aims at improving the collaboration amongst developers and overall improving the quality of software related practices and products used globally.

\begin{acks} 
	
	This work has been partially supported by the EMELIOT national research project, which has been funded by the MUR under the PRIN 2020 program (Contract 2020W3A5FY). 
	The work has been also partially supported by the European Union--NextGenerationEU through the Italian Ministry of University and Research, Projects PRIN 2022 PNRR \emph{``FRINGE: context-aware FaiRness engineerING in complex software systEms''} grant n. P2022553SL. We acknowledge the Italian ``PRIN 2022'' project TRex-SE: \emph{``Trustworthy Recommenders for Software Engineers,''} grant n. 2022LKJWHC.
\end{acks}

\bibliographystyle{ACM-Reference-Format}
\bibliography{main}


\end{document}